\documentclass[twocolumn]{cinc}
\usepackage{mathtools, graphicx, xurl, hyperref}
\newcommand{\comment}[1]{}

\title{Convolution-Free Waveform Transformers for Multi-Lead ECG Classification}
\author{Annamalai Natarajan\textsuperscript{1},
Gregory Boverman\textsuperscript{1},
Yale Chang\textsuperscript{1},
Corneliu Antonescu\textsuperscript{2},
Jonathan Rubin\textsuperscript{1}\\ \ \\
\textsuperscript{1}Philips Research North America, Cambridge MA, USA  \\
\textsuperscript{2}University of Arizona Banner Health, Phoenix AZ, USA}

\begin{document}
\maketitle


\begin{abstract}
We present our entry to the 2021 PhysioNet/CinC challenge --- a waveform transformer model to detect cardiac abnormalities from ECG recordings.
We compare the performance of the waveform transformer model on different ECG-lead subsets using approximately 88,000 ECG recordings from six datasets.
In the official rankings, team prna ranked between 9 and 15 on 12, 6, 4, 3 and 2-lead sets respectively. Our waveform transformer model achieved an average challenge metric of 0.47 on the held-out test set across all ECG-lead subsets. Our combined performance across all leads placed us at rank 11 out of 39 officially ranking teams.
\end{abstract}


\section{Introduction}

Cardiovascular diseases (CVDs) are the leading cause of death globally with an estimated 32\% of deaths worldwide in 2019 \cite{who}. It is important to detect cardiovascular diseases early so treatments can be provided to mitigate complications. The standard 12-lead ECG has been a popular choice in the diagnosis of various cardiac abnormalities, but,  more recently subsets of ECG leads have been used due to their size, cost, performance and ease of use. In this challenge, we utilize ECG recordings from subsets of standard 12-lead ECG to evaluate the efficacy of detecting cardiac abnormalities. The subsets of ECG leads include 12, 6, 4, 3, or 2 leads respectively. The challenge provides $\sim$88K ECG recordings from five training data sources assigned to belong to one or more of the 30 cardiac abnormalities. More details about this challenge can be found in \cite{2020ChallengePMEA, 2021ChallengeCinC}.

Prior work has shown deep neural networks to be successful in detecting cardiac abnormalities from 12-lead ECG signals \cite{natarajan2020wide}. In this work, we experiment with a waveform transformer model which differentially weights different parts of the inputs using an attention mechanism. This feature is a natural fit to this problem since cardiac abnormalities tend to be transient in ECG recordings.
Our waveform transformer model is similar to our entry in the 2020 PhysioNet challenge \cite{natarajan2020wide} with some modifications:
\begin{enumerate}
\item We remove the embedding network that applied a series of convolution operations to the input ECG recording and instead rely on a convolution-free approach where the input signal is split into smaller segments and fed directly to a transformer model.
\item We seed the network with pretrained vision transformer network weights that have been trained on two-dimensional still images. We tailor our architecture to allow single-dimensional waveform inputs to make use of these pretrained weights.
\end{enumerate}
We continue to train wide and deep networks that include 22 static ECG and demographic features, which are concatenated to the learned deep features of the network. These 22 features are similar to the ones listed in Table 1 in \cite{natarajan2020wide}.


\begin{figure*}
\begin{center}
    \fbox{\includegraphics[width=0.8\linewidth]{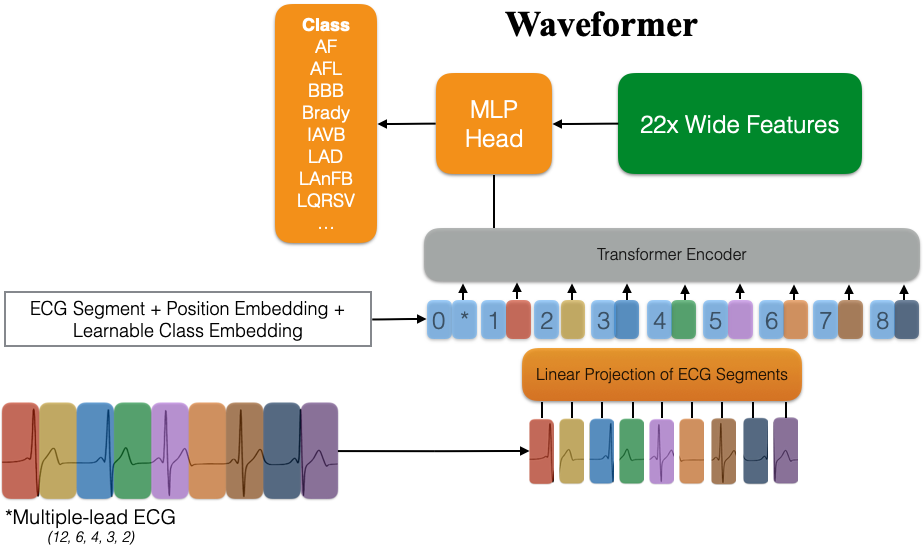}}
\end{center}
  \caption{Architecture diagram illustrating the Waveform Transformer (Waveformer) for multi-lead, multi-label ECG classification. Linear projection is used to allow matching of pretrained weights. Wide (static) features can be directly fed as input to the MLP head. Based upon an original image in \protect\cite{dosovitskiy2020image}.}
\label{fig:arch}
\end{figure*}


\section{Methods}

\subsection{Pre-processing}

Recordings from all databases provided by the challenge organizers \cite{2021ChallengeCinC, 2020ChallengePMEA}, including CPSC \cite{CPSC}, INCART \cite{INCART}, PTB \cite{PTB}, PTB-XL \cite{PTB-XL}, Chapman-Shaoxing \cite{Chapman-Shaoxing} and Ningbo \cite{Ningbo}, were used for model training. We first split the data into 10 folds utilizing multi-label stratification \cite{2017arXiv170201460S}. Each recording was standardized to a sampling rate of 500Hz. We apply a finite impulse response bandpass filter with bandwidth between 3 - 45 Hz. Each recording is also normalized so that each channels’ signal lies within the range of -1 to +1. We extract random fixed width windows from each recording across the subset of leads. We set the fixed width to be 7680 samples ($T=15.36$ seconds), which allows the signal to be split into divisible segments sizes. We apply zero-padding to the recordings at the end when the sequence length is less than $T$ seconds.

\begin{table}[ht!]
\centering
\begin{tabular}{|l|c|}
\hline
\textbf{Hyper-Parameter} & \textbf{Value}\\
\hline
\multicolumn{2}{|c|}{\textbf{Global}}\\
\hline
ECG window size (secs), $T$              & 15.36\\
Sampling frequency (Hz)             & 500\\ 
Batch size (train)                  & 128\\
Batch size (validation)             & 64\\
Wide feature size, $d_{wide}$       & 22\\
Deep feature size, $d_{deep}$       & 64\\
Number of classes, $d_{class}$      & 26\\
\hline
\multicolumn{2}{|c|}{\textbf{Waveform Transformer}}\\
\hline
ECG patch size, $d_{patch}$         & 64\\
Number of encoding layers           & 12\\
Embedding size, $d_{model}$         & 768\\
Number of heads                     & 12\\
Dimension of feed forward layer     & 768\\
Dropout                             & 0.1\\
\hline
\multicolumn{2}{|c|}{\textbf{Fully connected layers}}\\
\hline
FC 1 size                           & 64\\
FC 2 size                           & 26\\
Dropout                             & 0.2\\\hline
\end{tabular}
\caption{A listing of hyper-parameters selected to train the wide and deep neural network model for classification of cardiac abnormalities.}
\label{tab:params}
\end{table}

\subsection{Waveform Transformer}

An overview of the waveform transformer architecture is shown in Figure \ref{fig:arch}. Input to the network is a multi-lead ECG recording (e.g. 12, 6, 4, 3, or 2 leads). The input ECG recording is first broken up into smaller contiguous segments. Each segment undergoes a linear projection to embed it into a one dimensional vector that captures information for that time point in the overall recording. Linear projection of segments can be handled via a multi-layer perceptron (MLP) or convolution\footnote{Making the approach \emph{almost} convolution-free.}. A sequence of embedded segments are then fed to the transformer encoder. Positional embedding is used to retain sequence order information. In addition, an extra learnable class token is fed to the transformer network that attends to all other tokens. The transformer model consists of 12 layers, using 12 attention heads and an embedding dimension of 768. We rely on pretrained weights of a vision transformer \cite{dosovitskiy2020image} trained on still image data, and as such, need to ensure the dimensions of the network match. We chose a base vision transformer model trained on 16x16 image patches with an embedding dimension of 768. To make a final prediction, the learnable class embedding is sent as input into an MLP head consisting of two linear layers. Static (wide) features are concatenated to the final linear layer of the network and a sigmoid operation is applied to make binary predictions about 26 classes\footnote{Equivalent classes are combined, reducing 30 classes to 26.}. Table \ref{tab:params} provides details on model architecture, settings and hyper-parameters used in our experiments. All models were trained using PyTorch using base models and pretrained weights from PyTorch Image Models \cite{rw2019timm}


\begin{table*}[!ht]
    \centering
    \begin{tabular}{|c|c|c|c|c|}\hline
        \textbf{Leads} & \textbf{Training Set}        & \textbf{Validation Set} & \textbf{Test Set} & \textbf{Official Ranking} \\\hline\hline
        12    & $0.675 \pm 0.023$ &       0.58 &  0.49 &     9 \\
         6    & $0.658 \pm 0.026$ &       0.55 &  0.49 &     9 \\
         4    & $0.669 \pm 0.024$ &       0.55 &  0.46 &     14 \\
         3    & $0.663 \pm 0.026$ &       0.54 &  0.47 &     11 \\
         2    & $0.661 \pm 0.019$ &       0.53 &  0.44 &     15 \\
         All leads & -- & 0.55 & 0.47 & 11 \\\hline
    \end{tabular}
    \caption{Challenge scores for our final selected entry using 10-fold cross validation on the public training set, repeated scoring on the hidden validation set, and one-time scoring on the hidden test set as well as the official ranking on the hidden test set.}
    \label{tab:cv}
\end{table*}

\comment{
\begin{table*}[!ht]
\centering
\begin{tabular}{cccccc}
\hline
\textbf{Fold} & \textbf{12-lead} & \textbf{6-lead} & \textbf{4-lead} & \textbf{3-lead} & \textbf{2-lead}\\\hline
\textbf{1}       & 0.635	& 0.608	& 0.630	& 0.619	& 0.626\\
\textbf{2}       & 0.642	& 0.632	& 0.644	& 0.622	& 0.634\\
\textbf{3}       & 0.660	& 0.644	& 0.645	& 0.645	& 0.649\\
\textbf{4}       & 0.674	& 0.645	& 0.660	& 0.667	& 0.658\\
\textbf{5}       & 0.677	& 0.657	& 0.667	& 0.666	& 0.656\\
\textbf{6}       & 0.689	& 0.676	& 0.686	& 0.670	& 0.677\\
\textbf{7}       & 0.694	& 0.678	& 0.690	& 0.673	& 0.676\\
\textbf{8}       & 0.699	& 0.683	& 0.687	& 0.683	& 0.680\\
\textbf{9}       & 0.691	& 0.684	& 0.697	& 0.693	& 0.674\\
\textbf{10}      & 0.692	& 0.678	& 0.689	& 0.689	& 0.675\\\hline
\textbf{Mean$\pm$s.d.} & \textbf{0.675$\pm$0.023}  & \textbf{0.658$\pm$0.026 } & \textbf{0.669$\pm$0.024} & \textbf{0.663$\pm$0.026} & \textbf{0.661$\pm$0.019}\\\hline
\textbf{Official held-}  & \textbf{0.49}	& \textbf{0.49}	& \textbf{0.46}	& \textbf{0.47}	& \textbf{0.44}\\
\textbf{out test set} & & & & & \\\hline
\end{tabular}
\caption{Challenge metrics for 10-fold nested cross validation and on the official held-out test set different ECG-lead subsets.}
\label{tab:cv}
\end{table*}
}

\section{Results and Conclusions}
In this section, we present results from our waveform transformer model. Our setup is a standard 10-fold nested cross validation. In each fold, we utilize data from the validation fold to learn probability thresholds and other hyper-parameters. In Table \ref{tab:cv}, we list the challenge metric on the test folds as well as on the official test set for different ECG-lead subsets.
We observe that overall there is a monotonic improvement in the scores from the 2-lead to the 12-lead on both the test folds and the official held-out test set.
Our scores on the official held-out test set for the 12, 6, 4, 3 and 2 leads are 0.49, 0.49, 0.46, 0.47, 0.44 respectively, which shows a 0.05 units improvement in the challenge metric between 2 and 12 leads. We also observe that there is a significant gap in performance ($\sim0.2$) between model performance on the publicly available train dataset and the held-out dataset which hints at potential over fitting to the train dataset.

We examined the AUROCs for each cardiac abnormality computed using the probabilities as output by the transformer model on a single test fold. We observed similar trends in AUROC scores across different subsets of ECG leads. Our waveform transformer model achieved an AUROC of $0.85$ and $0.77$ on detecting low qrs voltage in 12-lead and 2-lead models respectively.
We hypothesize that low qrs amplitude ECG recordings are very similar to normal sinus rhythms but with low amplitudes which makes detecting them challenging.
We hypothesize that this poor performance on these cardiac abnormalities is that they often co-occur with other abnormalities that can be detected with high confidence.
This low AUROC was followed by detection of T wave abnormal, Q wave abnormal and T wave inversion with AUROC's in the range of $0.88$ to $0.90$ respectively.
All ECG lead subsets did exceptionally well in detecting pacing rhythm, left/right bundle block branch, tachycardia and bradycardia with AUROC's in the $0.98$ to $0.99$ range. This superior performance can be attributed to the unique physiological signatures embedded in ECG recordings and our inclusion of wide features such are heart rate.
The AUROC for normal sinus rhythm is $\sim 0.97$ across all ECG lead subsets.

\begin{figure*}
\begin{center}
    \fbox{\includegraphics[width=0.99\linewidth,trim=0 0.5cm 4cm 0,clip]{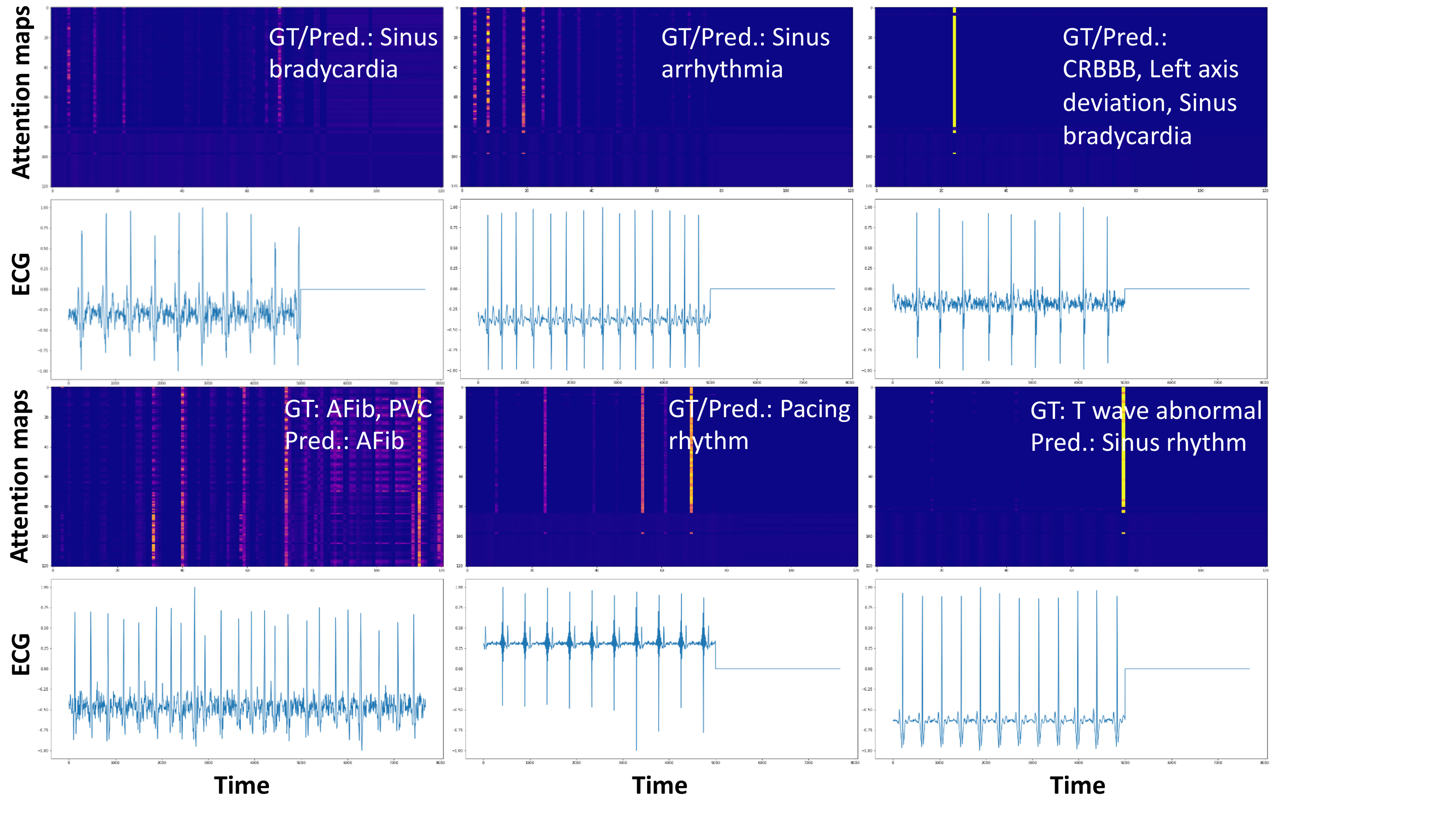}}
\end{center}
  \caption{Six randomly selected attention maps that show where attention is being paid in the corresponding ECG recordings (shown below). Recordings consist of 7680 samples that are encoded via a waveform transformer network into $N=120$ sequential embeddings. Attention heat maps are shown for lead II in the 12-lead setup. Also shown are the ground truth (GT) and predicted (Pred.) class labels.}
\label{fig:attn}
\end{figure*}

Lastly, we present the attention maps from the waveform transformer model in Figure \ref{fig:attn}. These attention maps illustrate which parts of the inputs are critical to making a final prediction much like a clinical expert manually scanning through ECG recordings. For each sample patient we show the ECG recording, attention maps along with ground truth (GT) and predicted (Pred.) labels. Both the x and y axis in the heatmaps represent time. Each cell represents a block of 64 samples (in time). Each sample (time point) can attend to all other time points. The high intensities (bright colors) along vertical time slices indicate that there are common time points that all time points are attending to. A potential shortcoming is highlighted for the bottom right patient which incorrectly predicts sinus rhythm, here the attention of all time points is on the length of the ECG recording which is non-informative.

\bibliographystyle{cinc}
\bibliography{references}

\begin{thebibliography}{10}
\expandafter\ifx\csname url\endcsname\relax
  \def\url#1{\texttt{#1}}\fi
\expandafter\ifx\csname urlprefix\endcsname\relax\def\urlprefix{URL }\fi

\bibitem{who}
{Cardiovascular Diseases}.
\newblock
  \url{https://www.who.int/en/news-room/fact-sheets/detail/cardiovascular-diseases-(cvds)}.
\newblock Accessed: 2021-08-10.

\bibitem{2020ChallengePMEA}
Perez~Alday EA, Gu A, Shah A, Robichaux C, Wong AKI, Liu C, et~al.
\newblock {Classification of 12-lead ECGs: the PhysioNet/Computing in
  Cardiology Challenge 2020}.
\newblock Physiological Measurement 2020;\hspace{0pt}41.

\bibitem{2021ChallengeCinC}
Reyna MA, Sadr N, Perez~Alday EA, Gu A, Shah A, Robichaux C, et~al.
\newblock {Will Two Do? Varying Dimensions in Electrocardiography: the
  PhysioNet/Computing in Cardiology Challenge 2021}.
\newblock Computing in Cardiology 2021;\hspace{0pt}48:1--4.

\bibitem{natarajan2020wide}
Natarajan A, Chang Y, Mariani S, Rahman A, Boverman G, Vij S, et~al.
\newblock A wide and deep transformer neural network for 12-lead ecg
  classification.
\newblock In 2020 Computing in Cardiology. IEEE, 2020;\hspace{0pt} 1--4.

\bibitem{dosovitskiy2020image}
Dosovitskiy A, Beyer L, Kolesnikov A, Weissenborn D, Zhai X, Unterthiner T,
  et~al.
\newblock An image is worth 16x16 words: Transformers for image recognition at
  scale.
\newblock arXiv preprint arXiv201011929 2020;\hspace{0pt}.

\bibitem{CPSC}
Liu F, Liu C, Zhao L, Zhang X, Wu X, Xu X, et~al.
\newblock {An Open Access Database for Evaluating the Algorithms of
  Electrocardiogram Rhythm and Morphology Abnormality Detection}.
\newblock Journal of Medical Imaging and Health Informatics
  2018;\hspace{0pt}8(7):1368--–1373.

\bibitem{INCART}
Tihonenko V, Khaustov A, Ivanov S, Rivin A, Yakushenko E.
\newblock {St Petersburg INCART 12-lead Arrhythmia Database}.
\newblock PhysioBank PhysioToolkit and PhysioNet 2008;\hspace{0pt}Doi:
  \url{10.13026/C2V88N}.

\bibitem{PTB}
Bousseljot R, Kreiseler D, Schnabel A.
\newblock {Nutzung der EKG-Signaldatenbank CARDIODAT der PTB \"uber das
  Internet}.
\newblock Biomedizinische Technik 1995;\hspace{0pt}40(S1):317--318.

\bibitem{PTB-XL}
Wagner P, Strodthoff N, Bousseljot RD, Kreiseler D, Lunze FI, Samek W, et~al.
\newblock {PTB-XL, a Large Publicly Available Electrocardiography Dataset}.
\newblock Scientific Data 2020;\hspace{0pt}7(1):1--15.

\bibitem{Chapman-Shaoxing}
Zheng J, Zhang J, Danioko S, Yao H, Guo H, Rakovski C.
\newblock {A 12-lead Electrocardiogram Database for Arrhythmia Research
  Covering More Than 10,000 Patients}.
\newblock Scientific Data 2020;\hspace{0pt}7(48):1--8.

\bibitem{Ningbo}
Zheng J, Cui H, Struppa D, Zhang J, Yacoub SM, El-Askary H, et~al.
\newblock {Optimal Multi-Stage Arrhythmia Classification Approach}.
\newblock Scientific Data 2020;\hspace{0pt}10(2898):1--17.

\bibitem{2017arXiv170201460S}
{Szyma{\'n}ski} P, {Kajdanowicz} T.
\newblock {A scikit-based Python environment for performing multi-label
  classification}.
\newblock ArXiv e prints February 2017;\hspace{0pt}.

\bibitem{rw2019timm}
Wightman R.
\newblock Pytorch image models.
\newblock \url{https://github.com/rwightman/pytorch-image-models}, 2019.

\end{thebibliography}

\begin{correspondence}
Jonathan Rubin\\
222 Jacobs St Cambridge, MA 02141, United States\\
jonathan.rubin@philips.com
\end{correspondence}

\balance

\end{document}